\documentclass[english]{sbrt}
\usepackage[english]{babel}
\usepackage[utf8]{inputenc}
\usepackage{newtxtext}

\usepackage{amsfonts}
\usepackage{graphicx} 
\usepackage{multirow}
\usepackage{amsmath}

\begin{document}

\title{Classification of Astronomical Spectra Using PCA-Compressed Flux and Inverse-Variance Features}

\author{Bruno Barreto and Marcio Eisencraft
\thanks{Departamento de Engenharia de Telecomunicações e Controle, Universidade de São Paulo, São Paulo-SP, e-mail: bruno.smbarreto@usp.br; marcioft@usp.br. This work was partially supported by CNPq (404081/2023-1).}%
}

\maketitle

\markboth{XLIV BRAZILIAN SYMPOSIUM ON TELECOMMUNICATIONS AND SIGNAL PROCESSING - SBrT 2026, SEPTEMBER 29TH TO OCTOBER 2ND, 2026, SALVADOR, BA}{}

\begin{abstract}
This paper evaluates a signal-processing and supervised-learning pipeline for classifying
SDSS DR17 astronomical spectra into stars, galaxies, and quasars.
Each spectrum is represented by its measured flux and inverse-variance
information, combining spectral shape with a wavelength-dependent  reliability profile. After resampling onto a common
logarithmic wavelength grid, the flux and inverse-variance vectors are  standardized and separately compressed using principal component analysis. The resulting components are concatenated and used to train several classifiers. The best
performance was obtained with the LightGBM gradient-boosting classifier, reaching 94.6\% accuracy and 92.1\%
balanced accuracy on the test set.
\end{abstract}

\section{Introduction}

Large-scale astronomical surveys have transformed observational
astrophysics into a data-intensive field, producing massive collections
of spectra from different types of astronomical objects. In particular, the Sloan
Digital Sky Survey (SDSS) has provided a large public database of
spectroscopic observations, enabling the development of automatic
methods for object identification and physical analysis
\cite{York_2000,ivezic2014statistics}. Since each spectrum can be interpreted as
a one-dimensional sampled signal containing continuum information,
absorption features, emission lines, noise, and instrumental effects,
spectral classification can be cast as a signal classification and pattern recognition problem.

The automatic classification of astronomical spectra into stars,
galaxies, and quasars is an important step in large survey pipelines \cite{Bolton2012}.
Stars are mostly characterized by stellar continua and absorption-line
patterns, galaxies often exhibit composite stellar populations, and quasars are typically associated
with broad emission lines and non-stellar continua \cite{Yip2004,VandenBerk2001}. These differences
make spectra rich sources of discriminative signal features. In this context, machine learning methods provide a
practical framework for learning nonlinear decision boundaries from high-dimensional spectral data.

In this work, we investigate supervised learning methods for
classifying SDSS spectra into three classes: STAR, GALAXY, and QSO (quasi-stellar object). The evaluated pipeline uses both spectral flux and
inverse-variance information, allowing the
classifiers to exploit the measured signal together with a wavelength-dependent reliability profile.  The spectra are resampled onto a common logarithmic
wavelength grid, standardized, compressed using principal component
analysis (PCA) \cite{Jolliffe2002-mg}, and then used as input to a set of
classification models ranging from linear methods, such as logistic regression and linear support
vector machines (SVMs) \cite{hastie_09_elements-of.statistical-learning,Cortes1995}, to ensemble-based gradient boosting approaches \cite{Friedman2001,Ke2017LightGBM}.

The remainder of the paper is organized as follows: Section~\ref{dataset}
describes the dataset, Section~\ref{dataprepro} presents the preprocessing
steps, Section~\ref{Model} details the classification models, Section~\ref{sec:results}
discusses the experimental results,  and Section~\ref{sec:conclusion} concludes the paper.

\section{Dataset}
\label{dataset}

The dataset used in this work was built from spectroscopic data in the SDSS Data Release 17 (SDSS DR17) catalog \cite{Abdurrouf_2022}. The corresponding FITS files \cite{Wells1981FITS} were retrieved for the selected objects. We retained only spectra from observations with acceptable quality (\texttt{PLATEQUALITY} = good or marginal), with no SDSS redshift/classification warning flags (\texttt{ZWARNING} = 0), and corresponding to the primary spectroscopic observation of each object (\texttt{SPECPRIMARY} = 1). For each spectrum, we extracted the measured flux, the logarithmic wavelength grid, the inverse variance provided by the SDSS pipeline, and the class label. The labels correspond to the three object classes considered in this work: STAR, GALAXY, and QSO.

After data selection, the final dataset
contained $32\,259$ spectra. The dataset was divided into training, validation, and test subsets using a stratified random split with proportions of 70\%, 15\%, and 15\%, respectively. This resulted in $22\,581$ spectra for training, $4\,839$ for validation, and $4\,839$ for testing.

The training set was imbalanced, with GALAXY spectra forming the majority class. It contained $14\,243$ GALAXY spectra, $4\,276$ STAR spectra, and $4\,062$ QSO spectra. This imbalance is relevant because a classifier trained without correction may become biased toward the most frequent class. Therefore, class weights inversely proportional to class frequencies were computed from the training set and used whenever supported by the classification model. This also motivated the use of balanced accuracy and macro-F1, in addition to overall accuracy, for performance evaluation.

\section{Data Preprocessing}
\label{dataprepro}

The raw SDSS DR17 spectra were preprocessed to obtain a common-grid representation suitable for supervised classification. We first determined the intersection of the log-wavelength ranges covered by the selected spectra, where \(\log\lambda=\log_{10}(\lambda)\) and  $\lambda$ is measured in angstroms. A fixed grid with $4\,000$
uniformly spaced points was then defined over this interval. The resulting grid spans approximately from \(\log\lambda=3.70\) to \(\log\lambda=3.80\), corresponding to about \(5\,000\)--\(6\,300\)~\AA.

The common wavelength grid was restricted to the interval shared by the selected spectra, providing a common representation while avoiding extrapolation.  Although some objects, especially QSOs, cover  broader wavelength ranges, using the union of all observed wavelength ranges  would require extrapolating or padding many STAR and GALAXY spectra outside their observed domains. This would introduce artificial signal components and could create nonphysical features in the input representation. 

To reduce the effect of absolute flux scaling, each spectrum was
standardized independently. Given a flux vector
\(\mathbf{x}_i \in \mathbb{R}^p\), where $p=4\,000$, the standardized spectrum
\(\tilde{\mathbf{x}}_i\) is defined by
\begin{equation}
\bar{x}_i=\frac{1}{p}\sum_{j=1}^{p}x_{ij},\quad
s_i=\sqrt{\frac{1}{p}\sum_{j=1}^{p}(x_{ij}-\bar{x}_i)^2},\quad
\tilde{x}_{ij}=\frac{x_{ij}-\bar{x}_i}{s_i}.
\end{equation}
The same per-spectrum standardization was applied to the inverse-variance vectors before dimensionality reduction. For the flux vectors, this standardization makes the model focus on
relative spectral shape and line patterns rather than on absolute flux scale. For the inverse-variance vectors, the standardized representation is interpreted as a relative wavelength-dependent reliability profile within each spectrum. 

\begin{figure}[hbt]
\begin{center}
\includegraphics[width=0.39\textwidth]{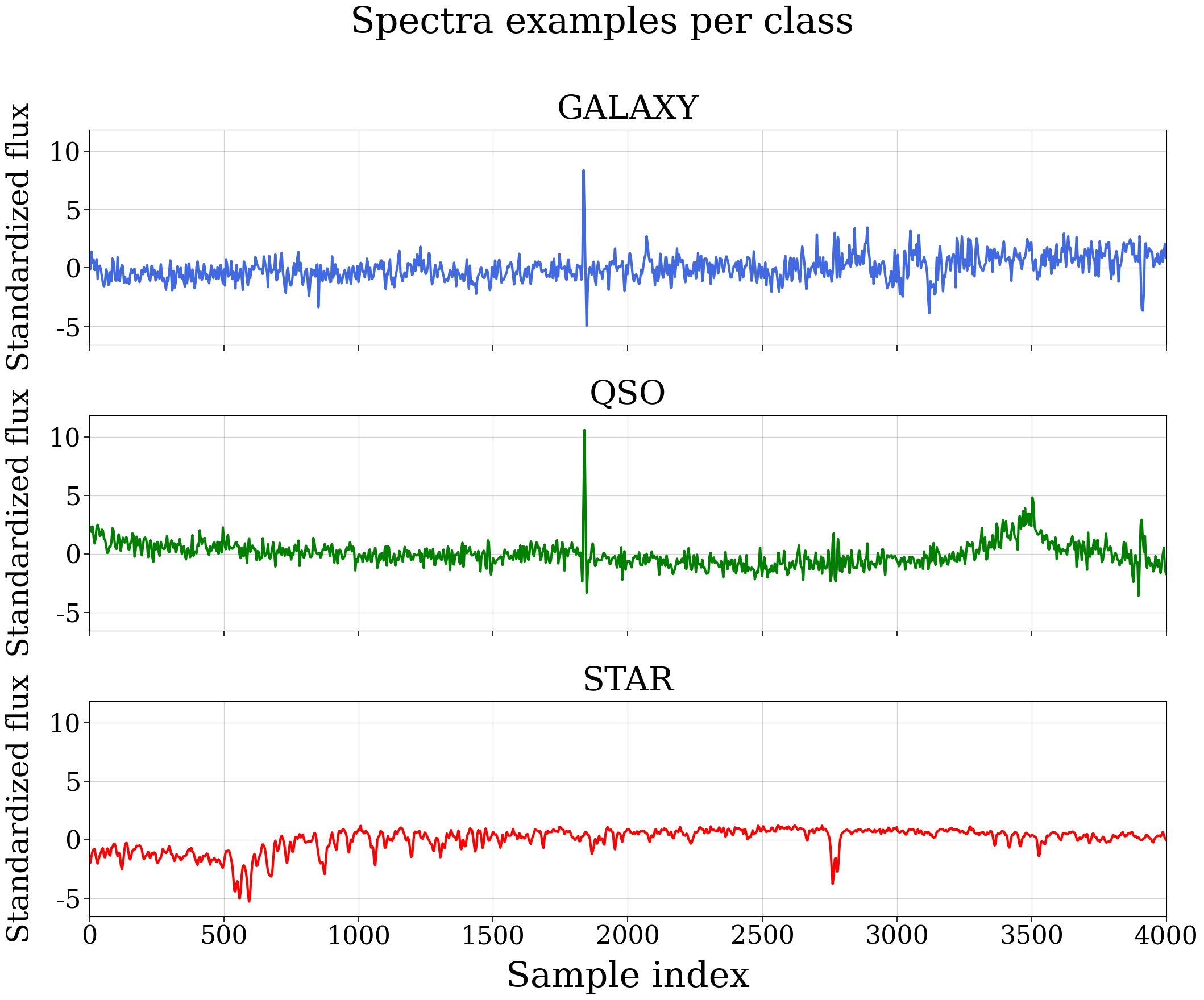}
\end{center}
\caption{Examples of resampled and per-spectrum standardized flux spectra
for GALAXY, QSO, and STAR objects. The common dimensionless scale emphasizes  relative spectral shapes
rather than absolute flux intensity.}
\label{fig:example}
\end{figure}

Figure~\ref{fig:example}
shows examples of resampled and standardized spectra from each class.

PCA was applied separately to
the standardized flux and inverse-variance matrices. In both cases, the PCA transformation
was fitted only on the training data and then applied to the validation
and test sets, avoiding information leakage. The first \(d=24\) principal components from each representation were retained, since the validation-set accuracy showed no substantial improvement beyond approximately 24 components. The resulting flux and inverse-variance components were concatenated and used as the input features for the classifiers.

\begin{figure}[hbt]
\begin{center}
\includegraphics[width=0.35\textwidth]{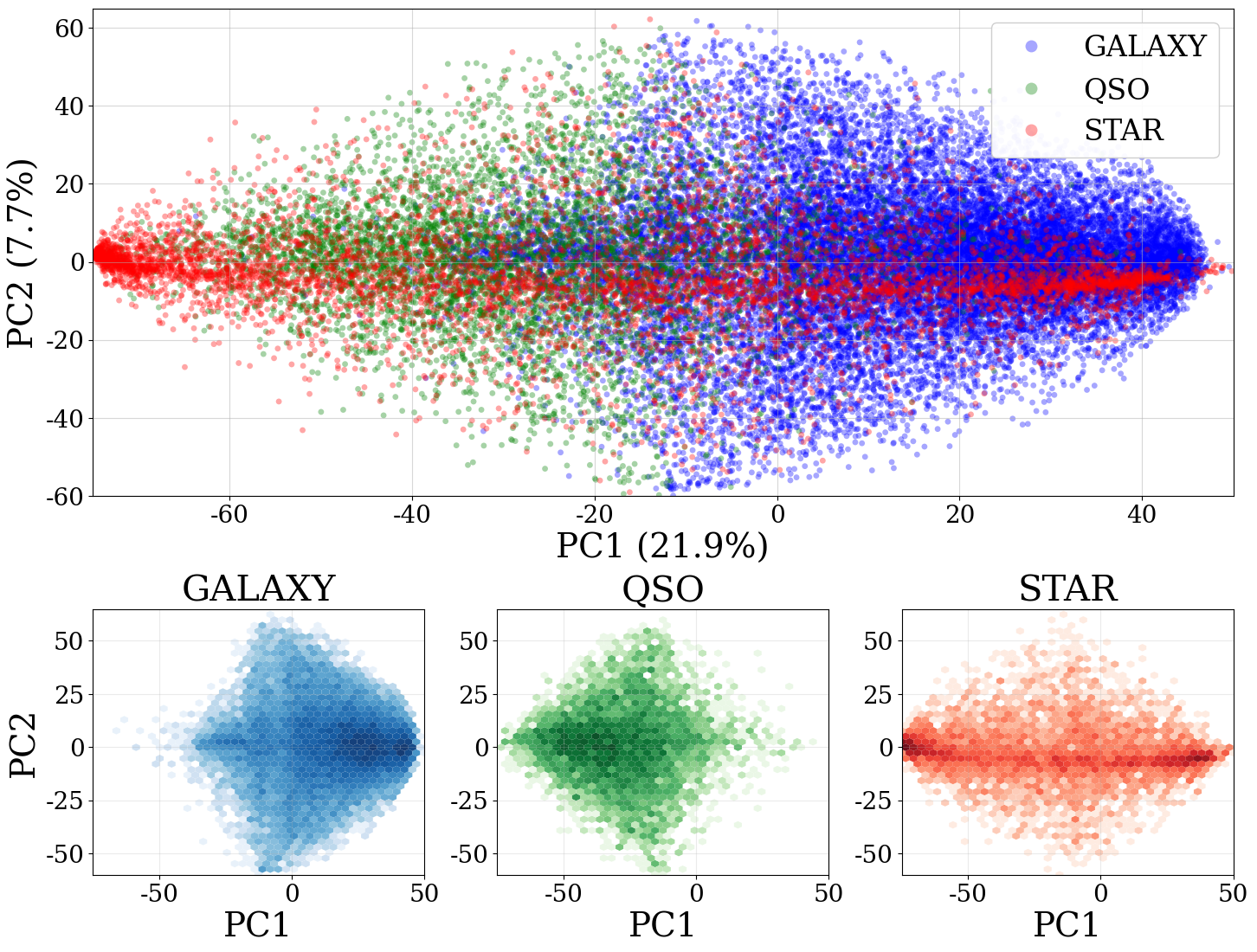}
\end{center}
\caption{Projection of the standardized flux spectra onto the first two
principal components, PC1 and PC2, which explain 21.9\% and 7.7\% of the variance, respectively. The upper panel shows the three classes jointly, while
the lower panels show the class-wise distributions for GALAXY, QSO, and STAR.}
\label{fig:dist_pca}
\end{figure}

Figure~\ref{fig:dist_pca} shows the projection of the
standardized flux spectra onto the first two principal components, which
explain 21.9\% and 7.7\% of the variance, respectively. The projection
reveals differences among the class distributions, but also substantial overlap. 
This indicates that the first two components alone are not sufficient for reliable class separation, motivating the use
of a higher-dimensional PCA representation followed by nonlinear classifiers.

\section{Classification Models}
\label{Model}

We consider the problem of learning a mapping from a preprocessed
spectrum to one of three astronomical classes: STAR, GALAXY, and QSO. For
the PCA-based classifiers, each object is represented by two feature
vectors: one obtained from the spectral flux and another from the inverse-variance profile. Let
\(\mathbf{c}^{(\mathrm{flux})}_i \in \mathbb{R}^{d}\) and \(\mathbf{c}^{(\mathrm{ivar})}_i \in \mathbb{R}^{d}\)
denote the corresponding PCA representations for the \(i\)-th spectrum. These vectors are
concatenated into 
\begin{equation}
\mathbf{z}_i =
\left[
\mathbf{c}^{(\mathrm{flux})}_i,
\mathbf{c}^{(\mathrm{ivar})}_i
\right]
\in \mathbb{R}^{2d}.
\end{equation}
The classifier then assigns \(\mathbf{z}_i\) to one of the three classes, possibly through class scores or posterior-probability estimates.

To evaluate the effect of model complexity on this classification task,
we benchmarked several supervised learning methods. First, we considered
linear classifiers, including multinomial logistic regression and a
linear support vector machine (SVM) \cite{hastie_09_elements-of.statistical-learning,Cortes1995}. These models provide simple
baselines for assessing whether the PCA-compressed representation can be effectively separated by linear decision boundaries.

We then evaluated generative and distance-based classifiers, including
Gaussian Naive Bayes (Gaussian NB), linear discriminant analysis (LDA), quadratic
discriminant analysis (QDA) \cite{hastie_09_elements-of.statistical-learning}, and $k$-nearest neighbors (k-NN) \cite{Cover1967}. These
models represent different assumptions about the feature space:
Gaussian NB assumes conditional independence among features, LDA and QDA
model class-conditional Gaussian distributions with different covariance
 structures, and k-NN performs classification based on local similarity in feature space.

Tree-based ensemble methods were also tested, including random forests \cite{random},
extremely randomized trees \cite{extreme}, gradient-boosting decision trees \cite{Friedman2001}, XGBoost \cite{xgboost} and
LightGBM \cite{Ke2017LightGBM}. These models can capture nonlinear interactions among the PCA
features through randomized ensembles or boosting.

Finally, a compact multilayer perceptron (MLP) \cite{Goodfellow-et-al-2016} was trained on the concatenated PCA representations as a nonlinear neural-network baseline. All models used the same training, validation, and test partitions. Due to the class imbalance in the training data, class weights were applied whenever supported by the corresponding algorithm. Hyperparameters were selected exclusively on the validation set, whereas the test set was held out for the final performance assessment. Performance was evaluated using accuracy (Acc.), balanced accuracy (Bal. Acc.), macro-averaged F1 score (Macro-F1), and weighted F1 score (W-F1), computed with scikit-learn \cite{Pedregosa2011}.

\section{Experimental Results}
\label{sec:results}

Table~\ref{tab:all_benchmark_results} summarizes the test-set
performance  using \(d=24\) PCA components per modality. The table compares classifiers trained using both flux
and inverse-variance information (F+I) against models trained only with
flux features (F). Overall, the best performance was obtained by the
gradient-boosting methods. In particular, LightGBM achieved the highest accuracy,
macro-F1, and weighted-F1 scores, reaching 0.946, 0.929, and 0.945,
respectively, while also obtaining a balanced accuracy of 0.921 in the F+I configuration. 

\begin{table}[htb]
\caption{Test-set performance with \(d=24\) PCA components per modality, using both flux and inverse-variance information (F+I) and only flux features (F).}
\label{tab:classification_results}
\centering
\scriptsize
\setlength{\tabcolsep}{2.8pt}
\begin{tabular}{lcccccccc}
\hline
\multirow{2}{*}{Model} 
& \multicolumn{2}{c}{Acc.} 
& \multicolumn{2}{c}{Bal. Acc.} 
& \multicolumn{2}{c}{Macro-F1} 
& \multicolumn{2}{c}{W-F1} \\
\cline{2-9}
& F+I & F & F+I & F & F+I & F & F+I & F \\
\hline
LightGBM      & \textbf{0.946} & 0.939 & \textbf{0.921} & 0.916 & \textbf{0.929} & 0.920 & \textbf{0.945} & 0.938 \\
Hist. GBDT    & 0.937 & 0.930 & \textbf{0.921} & 0.916 & 0.919 & 0.910 & 0.937 & 0.930 \\
XGBoost       & 0.945 & 0.938 & 0.912 & 0.906 & 0.927 & 0.919 & 0.944 & 0.938 \\
Small MLP     & 0.934 & 0.938 & 0.906 & 0.912 & 0.914 & 0.919 & 0.933 & 0.937 \\
Random Forest & 0.916 & 0.919 & 0.860 & 0.863 & 0.890 & 0.894 & 0.914 & 0.916 \\
QDA           & 0.903 & 0.883 & 0.851 & 0.835 & 0.872 & 0.850 & 0.901 & 0.881 \\
Logistic Reg. & 0.841 & 0.818 & 0.829 & 0.808 & 0.803 & 0.779 & 0.849 & 0.830 \\
Linear SVM    & 0.867 & 0.844 & 0.825 & 0.796 & 0.819 & 0.789 & 0.869 & 0.848 \\
Extra Trees   & 0.901 & 0.912 & 0.824 & 0.844 & 0.868 & 0.883 & 0.897 & 0.908 \\
LDA           & 0.855 & 0.837 & 0.782 & 0.747 & 0.797 & 0.767 & 0.854 & 0.835 \\
Gaussian NB   & 0.842 & 0.871 & 0.771 & 0.809 & 0.796 & 0.828 & 0.838 & 0.868 \\
k-NN          & 0.850 & 0.870 & 0.743 & 0.769 & 0.788 & 0.816 & 0.837 & 0.860 \\
\hline
\end{tabular}
\label{tab:all_benchmark_results}
\end{table}

The results indicate that the PCA-compressed representation preserves class-discriminative information despite the strong dimensionality reduction. Comparing the F+I and F configurations shows that
inverse-variance information improves the performance of the
best-performing gradient-boosting models. Although some classifiers performed  slightly better using only
flux features, the  results  suggest that inverse-variance
features can provide complementary information related to measurement
reliability and spectral uncertainty.

A confusion analysis of the best model, LightGBM with F+I features, indicates that most errors are concentrated in
the QSO class. The classifier correctly identifies 97.6\% of GALAXY spectra,
92.2\% of STAR spectra, and 86.5\% of QSO spectra. The dominant confusion
occurs when QSO spectra are classified as GALAXY, corresponding to 11.5\%
of the QSO test examples.

\section{Conclusion}
\label{sec:conclusion}

This work evaluated a signal-processing and supervised-learning pipeline for classifying SDSS DR17
spectra into STAR, GALAXY, and QSO classes using flux and
inverse-variance information. The PCA-compressed representation preserved discriminative spectral information  despite the dimensionality reduction, with
gradient-boosting methods achieving the strongest overall performance. In particular,
the LightGBM gradient-boosting classifier reached 94.6\% accuracy and 92.1\% balanced accuracy on the test set. The confusion analysis indicated that most remaining errors were associated with QSO spectra being classified as GALAXY. Overall, the results suggest that combining spectral preprocessing, reliability-related information, and nonlinear classifiers is a promising approach for astronomical spectral classification.

\bibliographystyle{ieeetr}
\bibliography{refe}

\end{document}